\documentclass[11pt,a4paper]{article}
\usepackage{tikz}
\usepackage{soul}
\usepackage{silence} 
\usepackage{jheppub}
\usepackage{graphicx}
\usepackage{orcidlink} 
\usepackage[makeroom]{cancel}
\usepackage{enumitem}
\usepackage{amssymb,amsmath,mathrsfs,bm,bbm,mathtools} 
\usepackage{amsthm}
\definecolor{ao(english)}{rgb}{0.0, 0.5, 0.0}
\hypersetup{colorlinks=true,linkcolor=blue,urlcolor=blue,citecolor=ao(english)}
\usepackage{stackrel}
\usepackage{multirow}
\usepackage[normalem]{ulem} 

\newtheorem*{theorem*}{Theorem}

\global\arraycolsep=1truept
\def\cob{\color{blue}}

\renewcommand{\leq}{\leqslant}
\renewcommand{\geq}{\geqslant}
\newcommand{\Eq}[1]{(\ref{#1})}
\newcommand{\Eqq}[1]{eq.~(\ref{#1})}

\newcommand{\be}{\begin{equation}}
\newcommand{\ee}{\end{equation}}
\newcommand{\ben}{\begin{equation*}}
\newcommand{\een}{\end{equation*}}
\newcommand{\ba}{\begin{eqnarray}}
\newcommand{\ea}{\end{eqnarray}}
\newcommand{\ban}{\begin{eqnarray*}}
\newcommand{\ean}{\end{eqnarray*}}
\newcommand{\bs}{\begin{subequations}}
\newcommand{\es}{\end{subequations}}
\newcommand{\bc}{\begin{center}}
\newcommand{\ec}{\end{center}}
\newcommand\nn{\nonumber\\}

\def\cL{\mathcal{L}}
\def\cM{\mathcal{M}}

\def\Re{\text{Re}}
\def\Im{\text{Im}}

\def\ve{\varepsilon}
\def\Om{\Omega}
\def\om{\omega}

\def\vp{\varphi}

\def\ke{k_\textsc{e}}
\def\xe{x_\textsc{e}}
\def\rme{{\rm e}}
\def\rmd{{\rm d}}
\def\rmi{{\rm i}}

\newcommand{\au}[2]{#1.~#2}
\newcommand{\arX}[1]{\href{http://arxiv.org/abs/#1}{{\cob arXiv:#1}}}
\newcommand{\oarX}[1]{\href{http://arxiv.org/abs/#1}{{\cob #1}}}

\newcommand{\book}[5]{\emph{#1}, #2, #3, #4 (#5)}
\newcommand{\books}[4]{\emph{#1}, #2, #3 (#4)}
\newcommand{\doin}[6]{\href{http://dx.doi.org/#1}{{\cob {\it #2 #3} {\bf #4} (#6) #5}}}
\newcommand{\doinn}[5]{\href{http://dx.doi.org/#1}{{\cob {\it #2} {\bf #3} (#5) #4}}}
\newcommand{\doij}[5]{\href{http://dx.doi.org/#1}{{\cob {\it #2} {\bf #3} (#5) #4}}}

\newcommand{\procm}[6]{in \emph{#1}, #2 (eds.), #3, #4, #5 (#6)}
\newcommand{\tia}[1]{\textit{#1},}

\begin{document}

\renewcommand{\thefootnote}{\fnsymbol{footnote}}
	
\title{Classicized dynamics and initial conditions in field theories with fakeons} 

\author[a,b]{Damiano Anselmi\,\orcidlink{0000-0001-6674-1328},}
\emailAdd{damiano.anselmi@unipi.it}
\affiliation[a]{Dipartimento di Fisica ``E.~Fermi'', Largo B.~Pontecorvo 3, 56127 Pisa, Italy}
\affiliation[b]{INFN, Sezione di Pisa, Largo B.~Pontecorvo 3, 56127 Pisa, Italy}

\author[c]{Gianluca Calcagni\,\orcidlink{0000-0003-2631-4588}}
\emailAdd{g.calcagni@csic.es}
\affiliation[c]{Instituto de Estructura de la Materia, CSIC, Serrano 121, 28006 Madrid, Spain}

\abstract{Theories with purely virtual particles (fakeons) do not possess a classical action in the strict sense, but rather a \emph{classicized} one, obtained by integrating out the fake particles at tree level. Although this procedure generates nonlocal interactions, we show that the resulting classicized equations of motion are not burdened with the need to specify infinitely many initial conditions. The reason is the inherent link between the fakeonic system and the parent higher-derivative local system: the solution space of the former is an appropriate subspace of solutions of the latter. A somewhat unexpected proviso is that, in order to avoid overcounting, the fakeon prescription must be obtained as a limit or special case of a more generic prescription. Ultimately, the set of degrees of freedom matches physical expectations, the extra ones (ghosts or otherwise) being removed by rendering them purely virtual. We illustrate the counting in simple linear solvable models and provide the general proof. Along similar lines, we analyze Dirac’s removal of runaway solutions in classical electrodynamics.}

	
\maketitle
\renewcommand{\thefootnote}{\arabic{footnote}}


\section{Introduction}

An option for building quantum field theories of gravity that are both renormalizable and unitary is to consider higher-derivative theories, which typically involve ghosts (i.e., propagators with poles multiplied by negative residues), and remove the unwanted degrees of freedom by inserting appropriate form factors. Over the years, this idea was pursued by Krasnikov, Kuz'min, Tomboulis, Modesto and others 
\cite{Krasnikov:1987yj,Kuzmin:1989sp,Tomboulis,Modesto:2011kw,Modesto,Briscese:2013lna,Modesto:2015lna,Modesto:2014lga,Modesto6,Calcagni:2023goc}. By using entire functions in the right places, the unphysical degrees of freedom are successfully eliminated from the spectrum, and unitarity holds \cite{Alebastrov:1972dxn,Briscese:2018oyx}. The drawback is that the resulting theory has a nonlocal Lagrangian. This entails an infinite degree of arbitrariness, and no physical principle is currently known to single out a unique theory that should describe Nature, although classes of theories are already able to yield predictions \cite{Modesto:2022asj,Calcagni:2022tuz}.

Starting with the pioneering work of Pais and Uhlenbeck \cite{Pais:1950za} and Efimov \cite{Efimov:1967pjn,Efimov:1973pjo,efimovnl3}, nonlocal quantum field theory has been widely investigated \cite{Modesto:2017sdr,Buoninfante:2022ild,BasiBeneito:2022wux,Koshelev:2023elc,Lanza}. 
It offers a rich and intriguing extension of conventional quantum field theory.
Nevertheless, before abandoning locality altogether, one may wish to explore less radical alternatives.
Within local quantum field theory, a strategy that has gained traction over the past decade is the one suggested long ago by Lee and Wick \cite{Lee:1969fy,Lee:1970iw,Lee:1969zze,Cutkosky:1969fq,Nakanishi:1971jj}, in which the ``abnormal particles'' are not truly removed from the spectrum, but decay quickly enough to remain unobservable in most applications. 
Related approaches involve propagators with complex poles \cite%
{Veltman:1963th,Yamamoto:1969vb,Yamamoto:1970gw,Nakanishi:1972pt,Mannheim:2018ljq,Buoninfante:2025klm,Liu:2022gun,Tokareva:2024sct,Asorey:2024mkb,BrCa}, analogies with QCD
\cite
{Holdom:2015kbf,deBrito:2023pli}, antilinear symmetries
\cite{Mannheim:2020ryw} and unstable ghosts \cite{Donoghue}.

A somewhat intermediate option is to introduce ``purely virtual particles" or fakeons (``particles'' that are always off the mass shell). Here one works within a ``parent" local quantum field theory, which generates the ``descendant'' fakeon theory through a certain projection.\footnote{Although no gauge symmetry is involved in the fakeon projection, a useful analogy is between a gauge-fixed theory and the theory where the gauge-trivial modes are projected out.} Because the unwanted degrees of freedom are truly removed from the spectrum, they must ultimately be integrated out. This yields a nonlocal quantum field theory of a special type. Notably: 1) the nonlocal sector is  restricted to the interaction sector, 2) renormalizability and unitarity hold, 3) the problem of infinite arbitrariness does not arise, thanks to the link between the descendant fakeon theory and the parent local theory. 

The concept of virtuality  lacks a classical analogue, prompting the question of how fakeons might manifest at the classical level. The purpose of this paper is to investigate aspects of the integrate-out procedure more closely and to clarify the relationship between the local parent theory and the nonlocal descendant theory in the classical limit. Among other points, we explain how the degrees of freedom are effectively lost in the descent. We also show that the initial conditions of the nonlocal field equations with fakeons are solely the physical ones. 


Fakeons offer a relatively simple resolution of the tension between the requirement of renormalizability and the violation of unitarity induced by higher-derivative operators. 
This goal can be achieved in four different but equivalent ways. The first two methods rely on the Anselmi–Piva (AP) or fakeon prescription, which can be implemented either 1) by combining the Lee--Wick prescription \cite{Lee:1969fy,Lee:1970iw,Lee:1969zze,Cutkosky:1969fq,Nakanishi:1971jj} for loop-energy integrals with a suitable deformation of the spatial-momentum integration domain \cite{Anselmi:2017yux,Anselmi:2017lia,Anselmi:2018kgz}, or 2) by averaging the analytically continued Euclidean amplitude around branch points involving the would-be fakeons \cite{Anselmi:2017yux,Anselmi:2017lia,Anselmi:2018kgz}. Alternatively, one may 3) introduce a new type of diagrammar \cite{Anselmi:2021hab},
or 4) employ specially defined non-time-ordered correlation functions \cite{Anselmi:2022qor}. 

The standard Cutkosky identities, which express the unitarity equation $S^{\dagger}S=1$ in diagrammatic form, can be  generalized to theories involving fakeons, for arbitrary diagrams \cite{Anselmi:2021hab,Anselmi:2022qor}.
In brief \cite{Anselmi:2021hab}, one first refines the  identities in theories containing only physical particles, recognizing that they decompose into a set of independent ``spectral optical identities,'' each holding at fixed spatial momenta and involving the same threshold. It then follows that fakeonizing a particle simply corresponds to replacing the standard Feynman diagrams with combinations of diagrams that $i$) eliminate the spectral optical identities associated with thresholds involving at least one fakeon, and $ii)$ leave all remaining identities unchanged.

It is worth to stress that 
the procedure of turning a degree of freedom into a fakeon, as well as the
nonlocal approach of Krasnikov \textit{et al.} \cite{Krasnikov:1987yj,Kuzmin:1989sp,Tomboulis,Modesto:2011kw,Modesto,Briscese:2013lna,Modesto:2015lna,Modesto:2014lga,Modesto6,Calcagni:2023goc}, apply to both physical particles and ghosts.\footnote{ 
In some sense, the Krasnikov \textit{et al.} idea of removing unwanted poles by suitable form factors can be viewed as a precursor of the concept of purely virtual particle, albeit at the nonlocal level.
Another line of inquiry into locality versus nonlocality is that of \cite{Anselmi:2024ocm}, which shows that, if a generic nonlocal theory admits a local limit, then that limit is a fakeon theory.}

Fakeons leave observable imprints of various types.
Differently from resonances, they exhibit a pair of bumps rather than a peak \cite{Anselmi:2023wjx}. 
In primordial cosmology, they can affect the spectra of scalar and tensor fluctuations in detectable ways \cite{Anselmi:2020lpp}. Moreover, they trigger a violation of microcausality \cite{Anselmi:2018bra}. 

As mentioned, another signature of fakeons appears in the classical dynamics, which differs from the usual one through nonlocal interaction terms. The correct equations of motion are not those derived from the \emph{interim} classical action (i.e., the higher-derivative action of the parent local theory), but those derived from the \emph{classicized} action, 
obtained through a process called \emph{classicization}. This process amounts to
collecting the tree diagrams that have physical particles on the external legs and purely virtual particles on the internal legs \cite{Anselmi:2018bra,Anselmi:2019rxg}. 
The key step is inverting a certain differential operator via the fakeon prescription. The outcome is that nonlocal interactions are generated in the resulting equations of motion.

Several aspects of the classicization deserve emphasis. First, the classicized equations of motion are approximate, because the correction terms are computed perturbatively in the couplings. Finding the fully resummed classicized Lagrangian can be highly nontrivial, but this is expected: purely virtual systems are intrinsically quantum and classicization carries typical quantum features (such as perturbativity) down to the classical limit.

Another puzzling aspect is that the classicized equations are nonlocal, which can obscure the physical meaning of the Cauchy problem for initial conditions \cite{Eliezer:1989cr,Moeller:2002vx}. Naively, solving nonlocal equations of motion seems impractical. In a sense, they involve infinitely many time derivatives, suggesting the need for infinitely many initial conditions.

We show that fakeonic nonlocal systems generated by classicization are immune to this difficulty, due to their link to the parent local higher-derivative system (the one associated with the interim action), which involves only finitely many time derivatives. The equations of the interim Lagrangian are recovered by acting on the classicized equations with a suitable differential operator. Hence, the solutions of the classicized equations form a subspace of those that solve the interim system, and involve only a finite number of arbitrary constants. 
We further show that the degrees of freedom of the fakeonic system match those of its expected physical subsector. 

We treat several solvable models and provide a general proof. We further demonstrate in detail how the unwanted initial conditions are eliminated in moving down from the parent local higher-derivative system to the descendant nonlocal system. 

The present investigation uncovers unexpected properties of classicization. For example, the number of initial conditions is reduced by the same amount regardless of the prescription adopted for the key operation, that is to say, the inversion of the fakeon operator. At the same time, in some particular cases it is not convenient to work directly with the fakeon prescription, because it may give the illusion that unwanted degrees of freedom survive. The correct counting is obtained by considering a generic prescription and reaching the fakeon one as a limit or particular case. 
We illustrate these features through several examples of linear classical-mechanics systems. 

A related, familiar case is Dirac’s removal of runaway solutions from the Abraham–Lo\-rentz force in classical electrodynamics. Dirac’s method \cite{Dirac} turns a local higher-derivative equation that needs three initial conditions into a nonlocal equation that needs only two. The arbitrary constant associated with the runaway solution is eliminated \cite{Anselmi:2018bra}. Again, the reduction works with any prescription, but only Dirac’s makes physical sense. Although Dirac’s prescription differs from the fakeon one (which does not apply to Dirac’s case), both trigger a violation of microcausality.

Additionally, we discuss the fakeon prescription in coordinate space. This is relevant to study the fakeon corrections to the equations of motion, which involve convolutions with the AP Green function $G_{\rm AP}(x^2)$. While in momentum space $\tilde{G}_{\rm AP}(k^2)$ is the average of the causal and anti-causal functions (which is also the average of the Feynman and anti-Feynman prescriptions), the structure of $G_{\rm AP}(x^2)$ in coordinate space is not equally simple. We show that it is a subtle combination of Wick- and counter-Wick-rotated complex modes. Yet, we fail to give a recipe to work out $G_{\rm AP}(x^2)$ directly in coordinate space, which confirms that the natural environment for fakeons is momentum space.

The structure of the paper is as follows. In section \ref{sec2}, we study the fakeon propagator $G_{\rm AP}(x^2)$ in coordinate space in three and four dimensions.  In section \ref{sec4}, we derive the classicized Lagrangian in a broad class of models. In section \ref{sec6}, we examine the degrees of freedom of the classicized equations of motion and their initial conditions. Section \ref{concl} contains the conclusions.

We work in Lorentzian signature $(+,-,\cdots,-)$. Momentum and coordinates in Euclidean signature are denoted by $\ke^{\mu}=(k^1,k^2,\cdots,k^D)^{\mu}$ and $\xe^{\mu}=(x^1,x^2,\cdots,x^D)^{\mu}$, respectively.


\section{Fakeon propagator in coordinate space}\label{sec2}

In this section, we consider typical fakeon propagators in coordinate space. 
The result is a combination of Wick- and counter-Wick-rotated complex contributions.

In the case of poles with complex conjugate masses, we start from the Euclidean propagator 
\begin{equation}
\tilde G_{\text{E}}(\ke^{2})=\frac{M^{2}}{(\ke^{2}+m^{2})^{2}+M^{4}}  \label{propa}
\end{equation}
in momentum space.
The fakeon prescription tells us that the propagator in Minkowski spacetime
is just (\ref{propa}) switched to Lorentzian signature (multiplied by a factor $-i$), i.e., 
\begin{equation}
\tilde G_{\text{AP}}(k^{2})=\frac{-i M^{2}}{(-k^{2}+m^{2})^{2}+M^{4}}.  \label{propaL}
\end{equation}%
Hence, we have
\begin{equation}
G_{\text{AP}}(x^{2})=\int\frac{\rmd ^{D}k}{(2\pi )^{D}}\frac{-i
M^{2}\rme ^{-ik\cdot x}}{(-k^{2}+m^{2})^{2}+M^{4}}  \label{L}
\end{equation}%
in coordinate space, where the integral is on Lorentzian momenta $k$.

We calculate $G_{\text{AP}}(x^{2})$ explicitly and show that it is not
related in a simple way to the Euclidean position-space propagator%
\begin{equation}
G_{\text{E}}(\xe^{2})=\int\frac{\rmd ^{D}\ke}{(2\pi )^{D}}\frac{M^{2}\rme ^{i\ke\cdot \xe}}{(k_{\text{E}}^{2}+m^{2})^{2}+M^{4}}.  \label{E}
\end{equation}%
In particular, $G_{\text{AP}}(x^{2})$ is neither the analytic continuation
\begin{equation}
G_{\text{L}}^{\text{an}}(x^{2})=G_{\text{E}}(-x^{2}+i\epsilon )  \label{L1}
\end{equation}%
of $G_{\text{E}}(x^{2})$ nor the average continuation (which is one way to formulate fakeons in momentum space \cite{Anselmi:2017yux,Anselmi:2017lia,Anselmi:2018kgz})
\begin{equation}
G_{\text{L}}^{\text{avg}}(x^{2})=\frac{1}{2}G_{\text{E}}(-x^{2}+i\epsilon )+%
\frac{1}{2}G_{\text{E}}(-x^{2}-i\epsilon ). \label{L2}
\end{equation}
Note that $G_{\text{AP}}(x^{2})$ is real, while $G_{\text{L}}^{\text{an}}(x^{2})$ is not. 


\subsection{Three dimensions}

In three spacetime dimensions, $D=3$, the Euclidean
propagator (\ref{E}) is%
\begin{equation}
G_{\text{E}}(\xe^{2})=\frac{i}{8\pi \sqrt{\xe^{2}}}%
\left( \rme ^{-m_{+}\sqrt{\xe^{2}}}-\rme ^{-m_{-}\sqrt{%
\xe^{2}}}\right) ,  \label{E3}
\end{equation}%
where $m_{\pm }\coloneqq \sqrt{m^{2}\pm iM^{2}}$.
On the other hand, the fakeon Green function (\ref{L}) reads
\begin{equation}
G_{\text{AP}}(x^{2})=-\frac{i}{8\pi }\left( \frac{\rme ^{-m_{+}\sqrt{%
-x^{2}-i\epsilon }}}{\sqrt{-x^{2}-i\epsilon }}+\frac{\rme ^{-m_{-}\sqrt{%
-x^{2}+i\epsilon }}}{\sqrt{-x^{2}+i\epsilon }}\right) .
\label{fg3}
\end{equation}
This result can be proved by evaluating (\ref{L}) in the two cases 
$x^{2}<0$, where we can choose $x^{0}=0$,
and $x^{2}>0$, where we can choose $\bm{x}=0$, and collecting the outcomes into a single expression. 

Neither (\ref{L1}) nor (\ref{L2}) coincides with (\ref{fg3}).
In particular, the difference between $G_{\text{AP}}(x^{2})$ at $x^{0}=0$ and the Euclidean Green function (\ref{E3}) at $x^{4}=0$ is due to the poles
located in the first and third quadrants of the complex $k^{0}$ plane. Ultimately, $G_{\text{AP}}(x^{2})$ is a subtle average between the two
contributions to $G_{\text{E}}(\xe^{2})$ with conjugate masses ``Wick
rotated'' in opposite ways.


\subsection{Four dimensions}

The Euclidean Green function in $D=4$ is evaluated in an analogous way.
Writing%
\begin{equation}
\frac{M^{2}}{(\ke^{2}+m^{2})^{2}+M^{4}}=\frac{i}{2}\left( \frac{1}{%
\ke^{2}+m_+^{2}}-\frac{1}{\ke^{2}+m_-^{2}}%
\right) ,
\label{Eu}
\end{equation}%
we see that the Fourier transform involves the usual Bessel function $K_{1}$
with masses $m_{\pm }$. Specifically,%
\begin{equation*}
G_{\text{E}}(\xe^{2})=\frac{i}{8\pi ^{2}\sqrt{\xe^{2}}}%
\left[ m_{+}K_{1}\left( m_{+}\sqrt{\xe^{2}}\right)
-m_{-}K_{1}\left( m_{-}\sqrt{\xe^{2}}\right) \right] .
\end{equation*}

As before, we can compute the fakeon Green function $G_{\text{AP}}(x^{2})$ by distinguishing the cases of timelike $x^{\mu}$ and
spacelike $x^{\mu}$, and collecting the results into a single expression. 
The outcome is
\begin{equation}
G_{\text{AP}}(x^{2})=-\frac{i}{8\pi ^{2}}\left[ \frac{m_{+}}{\sqrt{%
-x^{2}-i\epsilon }}K_{1}\left( m_{+}\sqrt{-x^{2}-i\epsilon }\right) +\frac{%
m_{-}}{\sqrt{-x^{2}+i\epsilon }}K_{1}\left( m_{-}\sqrt{-x^{2}+i\epsilon }%
\right) \right] .
\label{Ll}
\end{equation}%
Again, $G_{\text{AP}}(x^{2})$ is the sum of the $m_{+}$- and $m_{-}$%
-contributions to $G_{\text{E}}(\ke^{2})$, Wick rotated\ in
opposite ways.

The low-energy expansion is easily obtained from (\ref%
{propaL}):%
\begin{equation*}
G_{\text{AP}}(x^{2})\simeq -\frac{iM^{2}}{m^{4}+M^{4}}\left( 1-\frac{2m^{2}}{%
m^{4}+M^{4}}\Box \right) \delta ^{4}(x)+\cdots\,.
\end{equation*}%
Instead, the expansion close to the light cone gives%
\begin{equation}
G_{\text{AP}}(x^{2})\simeq \frac{i}{8\pi ^{2}}\mathcal{P}\frac{1}{x^{2}},
\label{HExpa}
\end{equation}%
where $\mathcal{P}$ denotes the Cauchy principal value. The Fourier
transform of this expression is $-i\pi\delta (k^{2})$. 

The high-energy limit of $M^{2}/[(k^{2}-m^{2})^{2}+M^{4}]$ can be studied by taking the masses to zero. Since $m^2$ is just a shift of $k^2$, we can keep $m>0$ at no cost. Because of the light cone, the result of the limit is not $M^{2}/(k^{2}-m^2)^{2}$. Instead, we have
\begin{equation}
\lim_{M\rightarrow 0}\frac{-iM^{2}}{(k^{2}-m^{2})^{2}+M^{4}}=-i\pi\delta
(k^{2}-m^{2})\,.\label{deltalim}
\end{equation}%
Note that this $\delta
(k^{2}-m^{2})$ does not signal the presence of a particle, because it contributes to the real part of the transition amplitude instead of the imaginary one.

\subsection{Two-derivative fakeons}\label{sec3}

When we fakeonize a standard particle with propagator%
\begin{equation}
\frac{i}{k^{2}-m^{2}+i\epsilon },
\label{nonHDcase}
\end{equation}%
we obtain the Cauchy principal value of $i/(k^2-m^2)$ in momentum space, and
\begin{equation}
G_{\text{AP}}^{\text{usual}}(x^{2})=-\frac{m}{8\pi ^{2}}\left[ \frac{%
K_{1}\left( m\sqrt{-x^{2}-i\epsilon }\right) }{\sqrt{-x^{2}-i\epsilon }}-%
\frac{K_{1}\left( m\sqrt{-x^{2}+i\epsilon }\right) }{\sqrt{-x^{2}+i\epsilon }%
}\right]
\label{poi}
\end{equation}%
in coordinate space \cite{Anselmi:2019nie}. Note that this expression
vanishes for spacelike $x$. The light-cone limit is 
\begin{equation}
G_{\text{AP}}^{\text{usual}}(x^{2})\simeq \frac{im}{4\pi }\delta (x^{2}),
\label{nonHD}
\end{equation}%
differently from (\ref{HExpa}). 

Multiplying (\ref{nonHDcase}) by an overall minus sign, we fakeonize a ghost, instead of a physical particle. We can also treat higher-derivative theories where (\ref{Eu}) is multiplied by a minus sign, since that sign does not change the nature of the problem we are dealing with.

Note that both higher-derivative and two-derivative fakeons violate Huygens' principle, since (\ref{Ll}) and (\ref{poi}) are nonzero away from the past light cone.


\section{Classicization}\label{sec4}

In this section, we derive the classicized Lagrangian and the classicized field equations in typical cases. In the first instance, we treat models where fakeons appear explicitly as independent fields. There, it is sufficient to integrate them out at the tree level with the AP prescription. Then we consider situations where the fakeons are implicit (e.g., hidden into higher-derivative kinetic terms) and must be first made explicit by introducing auxiliary fields, after which they can be integrated out.

Note that the classicized action is different from the more familiar one-loop effective action. The classicized action is obtained by integrating out only the fakeons (treating the other fields as external) and just keeping the tree diagrams, while the one-loop effective action is obtained by integrating on all the fields up to the one-loop diagrams. A possible source of concern is the infrared problems arising in the classical limit when massless particles are integrated out. The reason why fakeons are not affected by this issue is that they must be massive, because the reciprocal $1/m_{\chi}$ of their mass $m_{\chi}$ measures the violation of (micro)causality. Massless (or too light) fakeons are unacceptable, since they trigger a violation of causality, contrary to evidence at our scales. In quadratic gravity, for example (see subsection \ref{3.3}), the classicization amounts to just integrating out the heavy spin-two gravifakeon $\chi_{\mu\nu}$ at the tree level, treating the graviton as an external field.


\subsection{Integrating out the fakeons at the tree level}

The simplest situation is where the fakeons are ``integration ready'', i.e., they appear as independent fields. Then we can integrate them out right away by means of the AP prescription. At the tree level, this amounts to keeping the tree diagrams that have physical fields on the external legs and fakeons on the internal legs. The resulting classicized action is the sum of a local part plus nonlocal interactions
due to the fake particles.

Let $\varphi $ denote a physical field with a generic
self-interaction and $\phi$ an extra field that we want to quantize as a fake particle. We consider a prototypic cubic cross-interaction such as the one of the Lagrangian
\begin{equation}  \label{inter}
\mathcal{L}=\frac{1}{2}(\partial _{\mu }\varphi )^{2}-\frac{m^{2}}{2}\varphi
^{2}-V(\varphi)-\frac{1}{2M^{2}}\phi\left[ (\Box +m^{2})^{2}+M^{4}\right]
\phi -\frac{g}{2}\phi\varphi^{2}.
\end{equation}

As said, the classicized Lagrangian $\mathcal{L}_{\text{cl}}$ is the collection of tree diagrams that have fakeons $\phi$ on the internal legs and physical fields $\varphi$ on the external legs. Given that (\ref{inter}) is quadratic in $\phi$, $\mathcal{L}_{\text{cl}}$ can be obtained straightforwardly.
For example, we can take the $\mathcal{L}$ field equations 
\ba
&& (\Box+m^2)\varphi+V^{\prime}+g\phi\varphi=0\,,\label{loc1}\\
&&\frac{1}{M^2}[(\Box^2+m^2)^2+M^4]\phi+\frac{g}{2}\varphi^2=0\,,\label{loc2}
\ea
solve the second one for $\phi$ (with the fakeon Green function) and insert the solution into $\mathcal{L}$ itself. 
The result is 
\begin{equation}  \label{milan}
\mathcal{L}_{\text{cl}}=\frac{1}{2}(\partial _{\mu }\varphi )^{2}-\frac{m^{2}%
}{2}\varphi^{2}-V(\varphi)+\frac{g^{2}}{8}\varphi ^{2}
\left. \frac{M^2}{(\Box+m^2)^2+M^4}\right|_{\textrm{f}}\varphi ^{2}\,,
\end{equation}
where the subscript "$\textrm{f}$" is there to remind us that the AP (fakeon) prescription must be used to invert the operator. The last term in (\ref{milan}) can be written more explicitly with the formulas derived in the previous section, although we leave it as it stands here.

Note that, in the limit $M\rightarrow 0$, formula (\ref{deltalim}) tells us that $%
\mathcal{L}_{\text{cl}}$  becomes%
\begin{equation}
\lim_{M\rightarrow 0}\mathcal{L}_{\text{cl}}=\frac{1}{2}(\partial _{\mu
}\varphi )^{2}-\frac{m^{2}}{2}\varphi ^{2}-V(\varphi)+\frac{\pi g^{2}}{8}\varphi ^{2}\delta (\Box
+m^{2})\varphi ^{2}.
\end{equation}
The peculiar self-interaction encoded in the last term can be handled diagrammatically by working at $M>0$ and taking the limit $M\rightarrow 0$ at the very end. Although we do not yet know what applications this might have, it seems noteworthy. 

We can generalize this result to gauge interactions. Assume that $\varphi $ and $\phi $ are charged fields coupled to quantum electrodynamics. Then the classicized Lagrangian contains nonlocal corrections such as%
\begin{equation*}
\frac{\pi g^{2}}{8}\bar{%
\varphi}^{2}\delta (D_{\mu }D^{\mu }+m^{2})\varphi ^{2},
\end{equation*}%
in the limit $M\rightarrow 0$, where $D_{\mu }$ denotes the gauge-covariant derivative. Again, we can treat this self-interaction
diagrammatically at $M>0$ (e.g., by working in Euclidean space and performing the average continuation to Minkowski spacetime, according to the AP prescription) and let $M$ tend to zero at the end.

Another interest of the limit $M\rightarrow 0$ is that it allows us to compare the analytic continuation, the AP prescription, and the direct Minkowski calculation (see \cite{Anselmi:2025uzj}) in a relatively simple way. Add the fakeon self-interaction
\begin{equation}  \label{inter2}
-\frac{\lambda}{3!}\phi^{3}
\end{equation}
to (\ref{inter})
and consider the bubble diagram with two circulating fakeons. For simplicity, we set $m=0$, so that we can use the formul\ae\ of \cite{Anselmi:2025uzj}. If we take the analytic continuation from the Euclidean framework to the Minkowskian one, we get the following contribution to the amplitude $\cal M$:
\begin{equation*}
\pm \frac{i\lambda^2}{64\pi}\theta (p^{2})\,,
\end{equation*}%
depending on whether we approach the positive real axis from above or from below. The non-zero imaginary part signals that the analytic continuation turns on a production rate for particles that are absent at tree level, thereby violating unitarity.

The AP prescription amounts to taking the average of the two analytic continuations. In so doing, we obtain zero. In particular, the imaginary part of $\cM$ vanishes, so that there is no production rate for unphysical particles, in agreement with unitarity. 

Finally, if we calculate the bubble diagram directly in Minkowski spacetime, the contribution to the amplitude is (keeping $m$ non-zero, which is not a burden in this context)
\begin{equation*}
-\frac{i\lambda^2\pi^2}{2}\int \frac{\rmd ^{4}k}{(2\pi )^{4}}\delta (k^{2}-m^{2})\delta
[(p-k)^{2}-m^{2}]=-\frac{i\lambda^2}{64\pi}\theta (p^{2}-4m^{2})\sqrt{1-\frac{%
4m^{2}}{p^{2}}}.
\end{equation*}%
The optical theorem is violated (check \cite{Anselmi:2025uzj} for more details) and we have the propagation of unphysical degrees of freedom.


\subsection{Classicization of higher-derivative Lagrangians with a source}\label{subs}

Now we consider a simple higher-derivative Lagrangian with an external source and show how to remove the higher derivatives by introducing extra fields explicitly. Then we quantize them as fakeons and derive the classicized Lagrangian. We work in quantum mechanics (path integral) for simplicitly. A more general treatment of higher-derivative theories is presented in the next subsection. 

The Lagrangian we start from is
\begin{equation*}
L_{\text{HD}}=\frac{\dot{q}^{2}}{2}+\frac{\dddot{q}^{2}}{2M^{4}}+qF+\Delta L,
\end{equation*}%
where $F$ is an external force (which plays the role of a source) and $\Delta L$ is a $q$-independent part (ineffective on the equations of motion), which will be fixed later to subtract an analogous term from the final outcome. In what follows, integrations by parts are done without notice.

With a standard procedure, we introduce an auxiliary field $Q$ and obtain
the equivalent Lagrangian 
\begin{equation*}
L^{\prime }=\frac{\dot{q}^{2}}{2}-\frac{Q^{2}}{2}+\frac{\dot{q}\ddot{Q}}{%
M^{2}}+qF+\Delta L\,.
\end{equation*}%
At this point, we shift $q$ and $Q$ by means of the redefinitions%
\begin{equation*}
q=\bm{q}-\frac{\bm{\dot{Q}}}{M^{2}},\qquad Q=\bm{Q}+\frac{M^{2}}{%
\frac{\rmd ^{4}}{\rmd t^{4}}+M^{4}}\dot{F},
\end{equation*}%
where the operator in front of $\dot{F}$ is formal for the moment. The
result of the shift is the equivalent Lagrangian%
\begin{equation}
L^{\prime \prime }=\frac{1}{2}\bm{\dot{q}}^{2}+\bm{q}\frac{M^{4}}{%
\frac{\rmd ^{4}}{\rmd t^{4}}+M^{4}}F-\frac{1}{2}\bm{Q}^{2}-\frac{1}{2}\frac{\bm{%
\ddot{Q}}^{2}}{M^{4}}.  \label{Lpp}
\end{equation}%
We have chosen 
\begin{equation}
\Delta L=\frac{1}{2}\dot{F}\frac{M^{4}}{\left( \frac{\rmd ^{4}}{\mathrm{%
d}t^{4}}+M^{4}\right) ^{2}}\dot{F}
\label{unne}
\end{equation}%
to make the terms quadratic in $F$ disappear in $L^{\prime \prime }$.

It is clear now that the solution of the $\bm{Q}$ equation of
motion%
\begin{equation*}
\left( \frac{\rmd ^{4}}{\rmd t^{4}}+M^{4}\right) \bm{Q}=0
\end{equation*}%
is $\bm{Q}=0$, provided the inverse of the operator acting on $\bm{Q}$ is
uniquely defined by some {\it prescription}, so that there remains no room for nontrivial initial conditions on $\bm{Q}$ and its derivatives. If the prescription is the AP one (see below), then $\bm{Q}$ is the fakeon.

The argument just outlined cannot provide a privileged way to invert the operator ${\textrm d}^4\!/{\textrm d}t^4+M^4$. 
The fakeon prescription, which emerges from the requirement of unitarity in quantum field theory, 
follows from comparing different orders of the loop expansion.
Here we take it from granted. Once the answer is known, we also know the meaning of the operator
sandwiched in between $\bm{q}$ and $F$ in formula (\ref{Lpp}), and the classicized Lagrangian $L_{\text{cl}}$ follows straightforwardly.

Plugging the solution $\bm{Q}=0$ into $L''$, we obtain
\begin{equation}
L_{\text{cl}}=\frac{1}{2}\bm{\dot{q}}^{2}+\bm{q}\left. \frac{M^{4}}{%
\frac{\rmd ^{4}}{\rmd t^{4}}+M^{4}}\right\vert _{\text{f}}F.
\label{classi}
\end{equation}

Summarizing, the fakeon prescription is what gives meaning to the operator%
\begin{equation*}
\left. \frac{1}{\frac{\rmd ^{4}}{\rmd t^{4}}+M^{4}}\right|_{\textrm f},
\end{equation*}%
which is $G_{\text{AP}}(t)$ in coordinate space, apart from a factor. Precisely,
\begin{eqnarray}
\left.\frac{1}{\frac{\rmd ^{4}}{\rmd t^{4}}+M^{4}}\right|_{\textrm f} &\rightarrow
&\int_{-\infty }^{+\infty }\frac{\rmd \omega }{2\pi }\frac{\rme %
^{-i\omega t}}{\omega ^{4}+M^{4}}=\frac{\rme ^{-M|t|/\sqrt{2}}}{2\sqrt{2}%
M^{3}}\left( \cos \frac{M|t|}{\sqrt{2}}+\sin \frac{M|t|}{\sqrt{2}}\right) ,\label{w}
\\
\left(\left.\frac{1}{\frac{\rmd ^{4}}{\rmd t^{4}}+M^{4}}\right|_{\textrm f} \right)^2
&\rightarrow &\int_{-\infty }^{+\infty }\frac{\rmd \omega }{2\pi }\frac{%
\rme ^{-i\omega t}}{(\omega ^{4}+M^{4})^{2}}=-\frac{\rmd }{4M^{3}%
\rmd M}\left( \ref{w}\right) ,\nonumber
\end{eqnarray}%
the square being used in (\ref{unne}).

The equations of motion of the
classicized Lagrangian read%
\begin{equation}
\ddot{\bm{q}}(t)=\frac{M}{2\sqrt{2}}\int_{-\infty }^{+\infty }\rmd t^{\prime
}\rme ^{-M|t-t^{\prime }|/\sqrt{2}}\left[ (1-i)\rme %
^{iM|t-t^{\prime }|/\sqrt{2}}+(1+i)\rme ^{-iM|t-t^{\prime }|/\sqrt{2}}%
\right] F(t^{\prime }).
\end{equation}%
For example, the uniformly accelerated motion ($F=$ constant)
is unmodified: $\ddot{q}(t)=F$. 

Formula (\ref{w}) shows that, at the practical level, the violation of chronological ordering and microcausality is restricted to time intervals of order $1/M$.

We emphasize that fakeon models do not possess a ``classical theory'' in the strict sense. There can only be a
\textquotedblleft classicized\textquotedblright\ theory, inherited by the quantum
sector. A purely
virtual particle, like $\bm{Q}$ above, has no direct link to the classical world, because its classical limit vanishes.

\subsection{Classicization of quadratic Lagrangians}
\label{3.3}

The example treated in subsection \ref{subs} can be generalized to a special class of quantum field theories, which is still broad enough for our purposes, including applications to quantum gravity. 

Consider fields $\varphi ^{i}$ described by a higher-derivative Lagrangian of the form
\begin{equation}
\mathcal{L}^{\prime }(\varphi )=\mathcal{L}(\varphi )-\frac{\sigma }{2}\frac{%
\delta \mathcal{L}}{\delta \varphi ^{i}}\mathcal{F}^{ji}\mathcal{(\varphi )F}%
^{jk}\mathcal{(\varphi )}\frac{\delta \mathcal{L}}{\delta \varphi ^{k}},
\label{Lp}
\end{equation}%
where $\mathcal{L}(\varphi )$
is an ordinary, lower-derivative local Lagrangian, 
$\mathcal{F}^{ij}\mathcal{(\varphi )}$ is a
perturbatively local operator, $\sigma =\pm 1$ and summation over repeated indices is understood. In particular, $\mathcal{F}^{ij}$ may
contain derivatives acting to the left and to the right. 

We assume that the fields $\varphi ^{i}$ are bosonic, but the result we
derive can be easily generalized to fermions. When a regularization is
needed, we work with the dimensional one \cite{Bollini:1972ui,Bollini:1972bi,tHooft:1972tcz,Cicuta:1972jf}.

We call the Lagrangians of the form (\ref{Lp}) ``quadratic'', in analogy with the terminology
``quadratic gravity,'' which is a
particular case (see below). We want to isolate the degrees of freedom associated with the higher derivatives, to decide which are going to be quantized as fakeons and which are going to be quantized as physical particles.

Consider the generating functional%
\begin{equation}
Z^{\prime }(J)=\int [\mathrm{d}\varphi ]\exp \left( i\int \mathcal{L}%
^{\prime }(\varphi )+i\int J^{i}\varphi ^{i}\right) .  \label{zj}
\end{equation}%
Inserting extra bosonic fields $\chi ^{i}$, we can write%
\begin{equation}
Z^{\prime }(J)=\int [\mathrm{d}\varphi][\mathrm{d}\chi ]\exp \left( i\int 
\mathcal{\tilde{L}}(\varphi ,\chi )+i\int J^{i}\varphi ^{i}\right), 
\label{zip}
\end{equation}%
where%
\begin{equation}
\mathcal{\tilde{L}}(\varphi ,\chi )=\mathcal{L}(\varphi )+\frac{\delta 
\mathcal{L}}{\delta \varphi ^{i}}\mathcal{F}^{ji}\chi ^{j}+\frac{\sigma }{2}%
\chi ^{i}\chi ^{i}.  \label{ltilde}
\end{equation}%
The proof of the equivalence between (\ref{zj}) and (\ref{zip}) is based on the translation 
\begin{equation*}
\tilde{\chi}^{i}=\chi ^{i}+\sigma \mathcal{F}^{ij}\frac{\delta \mathcal{L}}{%
\delta \varphi ^{j}},
\end{equation*}%
which has Jacobian determinant equal to one and turns the Lagrangian (\ref%
{ltilde}) into 
\begin{equation*}
\mathcal{\tilde{L}}=\mathcal{L}^{\prime }(\varphi )+\frac{\sigma }{2}\tilde{%
\chi}^{i}\tilde{\chi}^{i}.
\end{equation*}%
Integrating on $\tilde{\chi}^{i}$, the right-hand side of (\ref{zip}) gives the right-hand side of (\ref{zj}).

Next, we mimick the procedure of subsection \ref{subs} and rearrange the
quadratic sector of $\mathcal{\tilde{L}}(\varphi ,\chi )$ by means of the
change of variables 
\begin{equation}
\varphi ^{i}=\hat{\varphi}^{i}-\mathcal{F}^{ji}(\hat{\varphi})\hat{\chi}%
^{j},\qquad \chi ^{i}=\hat{\chi}^{i}+\mathcal{A}_{0}^{ij}J^{j},  \label{chv}
\end{equation}%
where $\mathcal{A}_{0}$ is the constant matrix $W_{0}^{-1}\mathcal{F}_{0}$,
while $\mathcal{F}_{0}^{ij}=\mathcal{F}^{ij}(0)$ and 
\begin{equation*}
W_{0}=\sigma -\mathcal{F}_{0}\mathcal{L}_{0}^{\prime \prime }\mathcal{F}%
_{0}^{\text{t}}-\mathcal{F}_{0}U_{0}^{\text{t}},\qquad (\mathcal{L}%
_{0}^{\prime \prime })^{ij}=\frac{\delta ^{2}\mathcal{L}}{\delta \varphi
^{i}\delta \varphi ^{j}}(0),\qquad U_{0}^{ij}=\frac{\delta \mathcal{F}^{ik}}{%
\delta \varphi ^{j}}(0)\frac{\delta \mathcal{L}}{\delta \varphi ^{k}}(0).
\end{equation*}%
The superscript \textquotedblleft t\textquotedblright\ denotes the
transpose. It is easy to check that, upon using dimensional
regularization, the Jacobian determinant of (\ref{chv}) is equal to one.

We find 
\begin{eqnarray}
\hat{Z}(J) &\coloneqq &Z^{\prime }(J)\exp \left( -\frac{i\sigma }{2}\int J^{%
\text{t}}\mathcal{A}_{0}^{\text{t}}\mathcal{A}_{0}J\right)   \notag \\
&=&\int [\mathrm{d}\hat{\varphi}\mathrm{d}\hat{\chi}]\exp \left( i\int 
\mathcal{\hat{L}}_{2}(\hat{\varphi},\hat{\chi})+\mathcal{O}_{3}+i\int J^{%
\text{t}}\mathcal{B}_{0}\hat{\varphi}+i\int J^{\text{t}}\mathcal{O}%
_{2}\right) ,  \label{zipp}
\end{eqnarray}%
where
\begin{equation}
\mathcal{\hat{L}}_{2}(\hat{\varphi},\hat{\chi})=\mathcal{L}(\hat{\varphi})+%
\frac{1}{2}\hat{\chi}^{\text{t}}\left( W_{0}-U_{0}\mathcal{F}_{0}^{\text{t}%
}\right) \hat{\chi}^{j},\hspace{0in}\qquad \mathcal{B}_{0}=1+\mathcal{F}%
_{0}^{\text{t}}(W_{0}^{-1})^{\text{t}}(\mathcal{F}_{0}\mathcal{L}%
_{0}^{\prime \prime }+U_{0}).  \label{lpext}
\end{equation}%
Moreover, $\mathcal{O}_{3}$ and $\mathcal{O}_{2}$ denote interaction terms
that are at least cubic and quadratic in the fields, respectively. The
interaction terms proportional to $J$ can be treated by means of further
changes of variables, but they do not change the quadratic sector.

In this way, we have isolated the fields $\hat{\chi}$ due to the higher derivatives. 
Note that there are no $\hat{\chi}$-dependent source terms in the noninteracting sector. 

The simplest option is to quantize all the fields $\hat{\chi}$ as fakeons. However, in some cases we may want to quantize a subset of them as physical particles, for which additional source terms must be included. Quadratic gravity provides an example of system that admits the second option (see below).

Assuming that all the $\hat{\chi}$ are quantized as fakeons, the classicized Lagrangian is obtained by  integrating them out, which generates nonlocal corrections in the interaction
sector. The $\hat{\chi}$ Green function is $(W_{0}-U_{0}\mathcal{F}_{0}^{%
\text{t}})^{-1}$, treated with the AP prescription. The $\hat{\varphi}$ field equation reads%
\begin{equation}
-\mathcal{L}_{0}^{\prime \prime }\hat{\varphi}=\mathcal{B}_{0}^{\text{t}}J+%
\text{interactions.}
\label{lbj}
\end{equation}

For example, the case of subsection \ref{subs} is 
\begin{equation*}
\sigma =-1,\qquad J\rightarrow F,\qquad \varphi \rightarrow q,\qquad 
\mathcal{L}(\varphi )\rightarrow \frac{\dot{q}^{2}}{2},\qquad \mathcal{F}%
_{0}=\frac{1}{M^{2}}\frac{\mathrm{d}}{\mathrm{d}t},\qquad \mathcal{O}_{2}=%
\mathcal{O}_{3}=0,
\end{equation*}%
hence we find%
\begin{equation*}
W_{0}=-\left( 1+\frac{1}{M^{4}}\frac{\mathrm{d}^{4}}{\mathrm{d}t^{4}}\right)
,\qquad \mathcal{B}_{0}=-\left. W_{0}^{-1}\right|_{\textrm{f}}.
\end{equation*}

Quadratic gravity can be treated along the same lines \cite{Anselmi:2018bra,Anselmi:2018tmf}. In particular, $\mathcal{L}(\varphi )$ is the
Hilbert term 
\begin{equation*}
-\frac{1}{16\pi G}\sqrt{-g}(R+2\Lambda )
\end{equation*}
and we have
\begin{equation*}
\frac{\delta \mathcal{L}}{\delta \varphi ^{i}}\rightarrow -\frac{1}{16\pi G}%
\sqrt{-g}\left( R^{\mu \nu }-\frac{g_{\mu \nu }}{2}R-\Lambda g^{\mu \nu
}\right) ,\quad \mathcal{F}\rightarrow \sqrt{-g}\left[ a(g^{\mu \rho
}g^{\nu \sigma }+g^{\mu \sigma }g^{\nu \rho })+bg^{\mu \nu }g^{\rho \sigma }%
\right] ,
\end{equation*}%
$a$ and $b$ being constants. 
Here the set of fields $\hat{\chi}$ collects a massive scalar $\phi$ and a spin-2 massive field $\chi_{\mu\nu}$. The quadratic Lagrangian of the latter is a covariantized Pauli--Fierz Lagrangian multiplied by the wrong sign. By unitarity, we have to quantize $\chi_{\mu\nu}$ as a fakeon. Instead, we are allowed to quantize $\phi$ as a physical particle (which may play the role of the inflaton), because its quadratic Lagrangian carries the correct sign. If we keep $\phi$ physical, we have to add a source term $J_{\phi}\phi$ for it.  
 
Other higher-derivative gravities can be treated similarly, as long as
their Lagrangian can be phrased in the quadratic form (\ref{Lp}). 

\subsection{Higher derivatives and fakeons}

Now we analyze a simple example of equations of the form (\ref{lbj}) in the free-field limit with a source $J$.

Consider the higher-derivative model 
\begin{equation}
\cL=-\frac{1}{2M^2}\varphi (\Box ^{2}+M^{4})(\Box +\mu^{2})\varphi +J\varphi .
\label{LL}
\end{equation}%
We can manipulate the classical equations%
\begin{equation*}
\frac{1}{M^2}(\Box ^{2}+M^{4})(\Box +\mu^{2})\varphi =J
\end{equation*}%
by fakeonizing the left operator. This leads to the equation of motion%
\begin{equation*}
(\Box +\mu^{2})\varphi (x)=\left. \frac{M^2}{\Box ^{2}+M^{4}}\right\vert _{\text{%
f}}J(x)=i\int \rmd ^{4}y\hspace{0.02in}G_{\text{AP}}[(x-y)^2]\hspace{0.01in}%
J(y)
\end{equation*}
of the classicized Lagrangian (\ref{classi}), which here reads
\begin{equation}
\cL_{\text{cl}}=-\frac{1}{2}\varphi (\Box +\mu^{2})\varphi +\varphi \left. \frac{M^2
}{\Box ^{2}+M^{4}}\right\vert _{\text{f}}J.  \label{LLcl}
\end{equation}%
The AP Green function $G_{\textrm{AP}}$ is the one of (\ref{L}), or (\ref{Ll}), at $m=0$.

Using (\ref{HExpa}), in the high-energy limit we find
\begin{equation*}
\Box \varphi (x)+\mu^{2}\varphi (x)=-\frac{1}{8\pi ^{2}}\mathcal{P}\int 
\rmd ^{4}y\hspace{0.01in}\frac{1}{(x-y)^{2}}J(y)\eqqcolon \langle J(x)
\rangle_{\textrm{f}}\,.
\end{equation*}
For example, an oscillating source $J(y)=J_{0}\rme ^{i\omega y^{0}}\delta ^{3}(%
\bm{y})$ gives
\begin{equation}
\langle J(x)
\rangle_{\textrm{f}}=\frac{J_{0}\rme ^{i\omega x^{0}}}{8\pi|\bm{x}|}\sin
(\omega |\bm{x}|)=-\frac{iJ_{0}}{16\pi \omega}\left[ \rme ^{i\omega (x^{0}+|\bm{x}|)}-\rme ^{i\omega
(x^{0}-|\bm{x}|)}\right] .
\label{HDcase}
\end{equation}%

If we fakeonize a standard particle, as done in subsection \ref{sec3}, we have, using formula (\ref{nonHD}),
\begin{equation}
\langle J(x)\rangle_{\textrm{f}}=i\int \rmd ^{4}y\hspace{0.02in}G^{\textrm{usual}}_{\text{AP}}[(x-y)^2]\hspace{0.01in}%
J(y)=-
\frac{mJ_{0}}{8\pi |\bm{x}|}\left[ \rme ^{i\omega (x^{0}+|\bm{x}|)}+%
\rme ^{i\omega (x^{0}-|\bm{x}|)}\right] .
\label{nHDcase}
\end{equation}%
The behaviors of (\ref{HDcase}) and (\ref{nHDcase}) are similar, signaling that HD fakeons and non-HD fakeons have qualitatively similar, although quantitatively different properties. In particular, in both cases the violation of microcausality mostly propagates along the light cones.


\section{Initial conditions and number of degrees of freedom}\label{sec6}

In this section, we inquire about the number of degrees of freedom propagated by the fakeon models. 

A feature of the classicized Lagrangian \Eq{milan} is that its nonlocal sector is contained into the interaction sector. Therefore, in a perturbative context the former does not affect the counting of degrees of freedom, which turn out to be those of the free-field limit $g=0$.

Let us set $V(\varphi)=0$, for simplicity, since the local potential 
does not affect the main point of our argument. Inserting the perturbative expansion%
\begin{equation*}
\varphi =\varphi _{0}+\sum_{n=1}^{\infty }g^{2n}\varphi _{n}
\end{equation*}%
into the field equation at $V(\vp)=0$
\begin{equation*}
\Box \varphi +m^{2}\varphi =\frac{g^{2}}{2}\varphi\left. \frac{M^{2}}{(\Box +m^{2})^{2}+M^{4}%
}\right|_\textrm{f}\varphi ^{2},
\end{equation*}%
we obtain the equations satisfied by each $\varphi _{n}$, which are
\begin{equation*}
\Box \varphi _{0}+m^{2}\varphi _{0}=0,\qquad \Box \varphi _{n}+m^{2}\varphi
_{n}=\frac{1}{2}\sum_{i+j+k=n-1}\varphi _{i}\left.\frac{M^{2}}{(\Box +m^{2})^{2}+M^{4}}\right|_\textrm{f}%
\varphi _{j}\varphi _{k}\,,\qquad n>0.
\end{equation*}%
We see that every $\varphi _{n}$ is determined by two initial conditions. The
arbitrariness of each $\varphi _{n}$, $n>0$, is absorbed into a redefinition
of the zeroth-order solution $\varphi_{0}$. 

One may wonder whether extra initial conditions are added non-perturbatively in $g$. Are infinitely many initial conditions necessary to solve the nonlocal equations of motion of the classicized Lagrangian? The answer is No, because the fakeon problem is inherently tied to a local problem in the way we describe below. Precisely, the number of degrees of freedom coincides with the one of the perturbative expansion just recalled.

We begin with some exactly solvable models in one dimension. Then we give the general proof.


\subsection{Solvable models}\label{sec6a}

In this subsection, we consider simple models to illustrate how
a higher-derivative local system can be converted into a nonlocal one, keeping only the physical degrees of freedom.

\subsubsection{Dirac's removal of runaway solutions}

Before treating models with fakeons, we illustrate the procedure in a perhaps more familiar context, that is to say, Dirac's treatment
of the Abraham--Lorentz force in classical electrodynamics \cite{Dirac} (see also \cite{Jackson,AnselmiDirac}). It is well-known that an accelerated particle radiates. Under certain assumptions, the effects of the emitted radiation on the motion of the particle can be approximately described by the higher-derivative equation
\begin{equation}
ma-m\tau \dot{a}=F,\qquad \tau =\frac{2e^{2}}{3mc^3},
\label{Dir}
\end{equation}%
where $a$ is the acceleration, $F$ is an external force and $\tau$ is a fixed time parameter that can be calculated from first principles.

If we formally write
\begin{equation}
ma =\frac{1}{1-\tau \frac{\textrm{d}}{\textrm{d}t}}F
\label{HDDirac}
\end{equation}
and invert the operator $1-\tau  \textrm{d}/\textrm{d}t$ by requiring that it tends to 1 for $\tau\rightarrow 0$ (since we want $F=ma$ when the electromagnetic interactions are switched off), 
 we obtain the nonlocal equation \cite{Jackson,AnselmiDirac} (see also eq.\ (20) of \cite{BasiBeneito:2022wux})
\begin{equation}
ma=\frac{1}{\tau}\int_{t}^{\infty }\rmd %
t^{\prime }\hspace{0.02in}\rme ^{\frac{t-t^{\prime }}{\tau} }F(t^{\prime })\,.
\label{Dira}
\end{equation}
In this way, the higher-derivative local system (\ref{HDDirac}), which has one runaway solution, is converted into an inequivalent nonlocal system, (\ref{Dira}),
which lacks the runaway solution (as we show below in detail) and is fixed by one initial condition less. Moreover, the latter violates microcausality, since the right-hand side requires knowledge of the external force in a ``little bit of future'' (specifically, an amount of future of order $\tau$) beyond the future we want to predict.

For definiteness, we modify Dirac's case (\ref{Dir})-(\ref{Dira}) by replacing the external force $F$ with a harmonic force $-m\omega^2 x$. The higher-derivative and nonlocal equations are then
\ba
&&\text{(HD):}\qquad \ddot{x}-\tau \dddot{x}=-\omega ^{2}x,\\
&&\text{(NL):}\qquad \ddot{x}=-\omega ^{2}\left. \frac{1}{1-\tau \frac{\rmd }{%
\rmd t}}\right\vert _{\text{Dirac}}x=-\frac{\omega ^{2}}{\tau }%
\int_{t}^{\infty }\rmd t^{\prime }\,\rme %
^{\frac{t-t^{\prime}}{\tau}}x(t^{\prime })\,.\label{ldsys}
\ea
Although the two problems are related, they are not equivalent, to the extent that they involve different numbers of degrees of freedom. Moreover, although the NL system is nonlocal, it can be solved uniquely once we know the position and velocity at some initial time.

The strategy is as follows.
The NL equations yield the HD equations
when the operator $1-\tau \rmd /\rmd t$ is applied to both sides of \Eq{ldsys}. Hence, every NL solution is also an HD solution. 
The HD solutions can be easily classified. Writing them explicitly, it is easy to check which ones solve the NL problem and which do not. The result is that the HD system is determined by three initial conditions, while the NL one is determined by two.

In particular, one HD solution must be discarded because it makes the integral appearing on the right-hand side of the NL equation divergent. The other two HD solutions are instead fine and must be kept. They are all the solutions of the NL problem.

In formul\ae, the most general solution of the HD equation is 
\begin{eqnarray*}
x_{\text{HD}}(t) &=&\sum_{n=1}^{3}c_{n}\rme^{\lambda _{n}t},\qquad
3\tau \lambda _{n}=1-\frac{1}{\sigma _{n}W}-\sigma _{n}W, \\
\sigma _{1} &=&(-1)^{1/3},\quad \sigma _{2}=(-1)^{-1/3},\quad \sigma _{3}=-1,
\\
W &=&\left( 1+\frac{27}{2}\omega ^{2}\tau ^{2}-\frac{\Upsilon}{2}\right)
^{1/3},\qquad \Upsilon =3\sqrt{3}\, \omega \tau \sqrt{4+27\omega ^{2}\tau ^{2}},
\end{eqnarray*}%
and depends on the three arbitrary constants $c_{n}$. It is easy to show
that $0<W\leq 1$, $W+W^{-1}\geq 2$, $\Re\,\lambda _{1,2}\leq 0$, $\Im\,\lambda_{3}=0$, $\lambda _{3}\geq 1/\tau $. The last inequality tells us that $c_{3}$ parametrizes the runaway solution.

When $x_{\text{HD}}(t)$ is inserted into the NL equation, its right-hand side is divergent,
unless $c_{3}$ is set to $0$. Thus, the most general solution of the NL equation is%
\begin{equation*}
x_{\text{NL}}(t)=c_{1}\rme ^{\lambda _{1}t}+c_{2}\rme ^{\lambda
_{2}t},
\end{equation*}%
and correctly depends on two arbitrary constants only. The runaway solution has disappeared.

Dirac's prescription for the integral appearing on the right-hand side of the NL
equation \Eq{ldsys} is not essential for reducing the number of initial conditions. We can move to the most general prescription by assigning a finite value $a$ to the integral’s upper limit. Then the solution of the NL system has the form%
\begin{equation*}
x_{\text{NL}}(t)=c_{1}\rme ^{\lambda _{1}t}+c_{2}\rme ^{\lambda
_{2}t}+c_{3}(c_{1},c_{2},a)\rme ^{\lambda _{3}t},
\end{equation*}%
where $c_{3}$ is not arbitrary but depends on the other two coefficients
and on $a$:
\ben
c_3=-(\lambda_3\tau-1)\textrm{e}^{-a\lambda_3}\left(\frac{c_1 \textrm{e}^{a\lambda_1}}{\lambda_1\tau-1}+\frac{c_2\textrm{e}^{a\lambda_2}}{\lambda_2\tau-1}\right).
\een
Fixing the prescription amounts to fixing $a$, thereby removing one
degree of arbitrariness. Since $\Re(\lambda _{1,2}-\lambda_3)< 0$, the function $c_3(c_{1},c_{2},a)$ tends to zero when $a$ is sent to $+\infty$ (for arbitrary $c_1$ and $c_2$), giving back Dirac's prescription in the limit.

Ultimately, we see that the number of initial conditions of the nonlocal equation is always two, with no need to advocate
the recovery of $F=ma$ for $\tau\rightarrow 0$. The only role of this requirement is
to enforce a particular prescription among the others.

Now we analyze solvable models involving fakeons. 

\subsubsection{Tachyonic fakeon two-derivative model}

We start from the problem
\ba
&&\text{(HD):}\qquad \ddot{x}-\tau ^{2}\ddddot{x}=-\omega ^{2}x\,,\label{both}\\
&&\text{(NL):}\qquad \ddot{x}=-\omega ^{2}\left. \frac{1}{1-\tau ^{2}\frac{\rmd %
^{2}}{\rmd t^{2}}}\right\vert _{\text{f}}x=-\frac{\omega ^{2}}{2\tau }%
\int_{-\infty }^{\infty }\rmd t^{\prime }\hspace{0.01in}\rme %
^{-\frac{|t-t^{\prime }|}{\tau}}x(t^{\prime })\,.\label{6.4}
\ea
It is easy to check that applying $1-\tau^2 \text{d}^2/\text{d}t^2$ to both sides of \Eq{6.4} one obtains \Eq{both}.

The HD system admits a solution with four arbitrary constants,
\begin{equation}
x_{\text{HD}}(t)=c_{1}\rme ^{\lambda _{1}t}+c_{2}\rme ^{\lambda
_{2}t}+c_{3}\rme ^{\lambda _{3}t}+c_{4}\rme ^{\lambda
_{4}t}.
\label{mostHD}
\end{equation}
Nevertheless, only two
solutions solve the NL system:%
\begin{equation}\label{sol2hand}
x_{\text{NL}}(t)=c_{1}\rme ^{i\Omega t}+c_{2}\rme ^{-i\Omega
t},\qquad \Omega =\frac{1}{\tau \sqrt{2}}\sqrt{\sqrt{1+4\tau ^{2}\omega ^{2}}%
-1},\qquad c_2=c_1^*.
\end{equation}%
The other two, $\rme ^{\lambda _{3}t}$ and $\rme %
^{\lambda _{4}t}$, have $\lambda_{3,4}$ real with $|\lambda _{3,4}|\geq 1/\tau $, hence
they make the integral appearing on the right-hand side of the NL equation \Eq{6.4} divergent. 

Note that $\Om\to 0$ in the limit $\om\to 0$, where the NL system reduces to $\ddot x=0$.
The constants $c_{1,2}$ are replaced by $(a_0\Omega\mp \rmi a_1)/(2\Omega)$, where $a_0$ and $a_1$ are real, so that $x_{\textrm{NL}}(t)\rightarrow a_0+a_1t$.

As before, it is interesting to extend the discussion to the most general case, where the fakeon Green function of (\ref{6.4}) is replaced by a generic inverse of the operator $1-\tau^2 \text{d}^2/\text{d}t^2$.
Then the NL equation reads
\begin{equation}
\text{(NL):}\qquad \ddot{x}=-\frac{\omega ^{2}}{2\tau }%
\int_{a }^{t}\rmd t^{\prime }\hspace{0.01in}\rme %
^{\frac{t^{\prime }-t}{\tau}}x(t^{\prime })-\frac{\omega ^{2}}{2\tau }%
\int_{t}^{b}\rmd t^{\prime }\hspace{0.01in}\rme %
^{\frac{t-t^{\prime }}{\tau}}x(t^{\prime })\,,
\label{NLprimo}
\end{equation}
where $a$ and $b$ are arbitrary. Again, one can check that applying $1-\tau^2 \text{d}^2/\text{d}t^2$ to the right-hand side of \Eq{NLprimo} gives $-\om^2 x(t)$. The rest proceeds as before: we insert the most general solution (\ref{mostHD}) into (\ref{NLprimo}) and check when the latter is satisfied. We find that $c_3$ and $c_4$ are not independent, but functions of $c_1$ and $c_2$, as well as of $a$ and $b$. 
For example, for $\omega$ small we have the relations
\begin{equation}
    c_3=\frac{1}{2}(c_1+c_2)\tau^2\omega^2\text{e}^{a/\tau}+{\cal O}(\omega^3),\qquad
    c_4=\frac{1}{2}(c_1+c_2)\tau^2\omega^2\text{e}^{-b/\tau}+{\cal O}(\omega^3).
\end{equation}
We do not report the expressions for generic $\omega$ because they are quite lengthy.

We conclude that the number of initial conditions is two, for any choice of $a$ and $b$. {\sl De facto}, the missing degrees of freedom $c_3$ and $c_4$ are transferred into the choice of prescription for the inverse of $1-\tau^2 \text{d}^2/\text{d}t^2$, which is encoded into $a$ and $b$. The fakeon case is $a=-\infty$, $b=\infty$, where $c_3$ and $c_4$ tend to zero.

Note that the integrals appearing on the right-hand side of (\ref{NLprimo}) are well-defined for every finite $a$ and $b$. This means that the divergence we found in the fakeon case was an artifact of the prescription. It is not crucial for the reduction of the number of degrees of freedom. 

Recall that the AP fakeon prescription follows from the requirement of unitarity, a property that cannot be leveraged at the classical level. Yet, in view of the general proof given below, it is important to stress that a generic prescription (irrespectively of whether it is unitary or not at the quantum level) is able to reduce the number of initial conditions properly. Often, working directly with the fakeon prescription may give the impression that more initial conditions survive than the expected ones. Reaching fakeons as limits from generic prescriptions settles the matter unambiguously. The following variant of the model just considered clarifies what we mean by this.

\subsubsection{Tachyonic fakeon two-derivative model with repulsive force}

To appreciate this point better, 
consider the same model but flip the sign of the force. We can study this variant by writing $\omega=i\nu$ in the formulas above and assuming $\omega<1/(2\tau)$. Everything proceeds as before, with the caveat that if we work directly in the fakeon case ($a=-\infty$, $b=+\infty$), no reduction 
$x_{\text{HD}}\rightarrow x_{\text{NL}}$ seems to take place, because the integral on the right-hand side of the NL equation is convergent for the most general $x_{\text{HD}}$. To find the correct number of initial conditions, we have to work at finite $a$ and $b$, where the reduction works properly, and later take the limits $a\rightarrow -\infty$, $b\rightarrow +\infty$. It turns out that such limits are regular and give $c_3=c_4=0$. 


What just said teaches us that if we want to ``fakeonize'' an operator, we have to work in a generic environment (i.e., use the most general Green function that inverts the operator), where the right number of initial conditions follows straightforwardly. Particular cases, such as the fakeon one, must be reached as limits.

\subsubsection{Standard fakeon two-derivative model}

More typical fakeons are the ones of the problem
\ba
&&\text{(HD):}\qquad \ddot{x}+\tau ^{2}\ddddot{x}=-\omega ^{2}x\,,\\ 
&&\text{(NL):}\qquad \ddot{x}=-\omega ^{2}\left. \frac{1}{1+\tau ^{2}\frac{\rmd %
^{2}}{\rmd t^{2}}}\right\vert _{\text{f}}x=-\frac{\omega ^{2}}{2\tau }%
\int_{-\infty }^{\infty }\rmd t^{\prime }\hspace{0.01in}\sin
\left(\frac{|t-t^{\prime }|}{\tau}\right) x(t^{\prime })\,.  \label{fHD}
\ea
The crucial difference with respect to the previous case is the oscillating
behavior of the fakeon Green function%
\begin{equation*}
\left. \frac{1}{1+\tau ^{2}\frac{\rmd ^{2}}{\rmd t^{2}}}%
\right\vert _{\text{f}}=\mathcal{P}\frac{1}{1+\tau ^{2}\frac{\rmd ^{2}}{%
\rmd t^{2}}}\rightarrow \frac{1}{2\tau }\sin\left(\frac{|t-t^{\prime }|}{\tau}\right) ,
\end{equation*}
which is the Fourier transform of the Cauchy principal value of $1/(1-\tau
^{2}e^{2})$, where $e$ denotes the energy. 

The most general solution $x_{\text{HD}}(t)$ of the HD equation has the form (\ref{mostHD}), with four arbitrary constants $c_i$, $i=1,\ldots 4$. The problem is that all functions (\ref{mostHD}) seem to solve the NL\ problem as well, as long as the oscillating
contributions to the integral at infinity are dropped. 

On the contrary, we expect that 
the NL equation admits only two solutions, identified by the $\lambda_i$'s that
vanish for $\omega \rightarrow 0$. Since their expressions contain $\sqrt{1-4\omega^2\tau^2}$ (see (\ref{survi})), different behaviors occur at $\omega<1/(2\tau)$ and $\omega>1/(2\tau)$. Being us interested in the region with $\omega=0$, we assume $\omega<1/(2\tau)$. 

As before, we can bypass the difficulty by reaching the fakeon prescription as a limit. The first attempt amounts to starting from the most general inverse of the operator $1+\tau^2 \textrm{d}^2/\text{d}t^2$.
We obtain the nonlocal NL system
\begin{equation}
\text{(NL):}\quad \ddot{x}=-\frac{\omega^2}{2\tau}\int_a^t
\sin\!\left(\frac{t-t^\prime}{\tau}\right)x(t^\prime)\,\textrm{d}t^\prime+\frac{\omega^2}{2\tau}\int_t^b
\sin\!\left(\frac{t-t^\prime}{\tau}\right)x(t^\prime)\,\textrm{d}t^\prime\,,
\label{mostNL}
\end{equation}
where $a$ and $b$ are finite and arbitrary. Clearly, the fakeon limit is $a\rightarrow-\infty$, $b\rightarrow+\infty$.

When $x_{\text{HD}}(t)$  is inserted into the NL equation (\ref{mostNL}), one verifies that the latter is satisfied only if the coefficients $c_i$ are related by two conditions, which leave two degrees of arbitrariness. Since this is true for arbitrary finite $a$ and $b$, one expects that it is also true in the fakeon limit. However, the fakeon limit involves oscillations that are hard to control unambiguously. This problem is originated by the Cauchy principal value, which is a distribution.

To remove this extra difficulty, we
represent the principal value as the limit%
\begin{equation*}
\mathrm{\lim_{\ve \rightarrow 0}}\frac{1+\tau ^{2}\frac{\rmd ^{2}}{%
\rmd t^{2}}}{\ve ^{2}+\left( 1+\tau ^{2}\frac{\rmd ^{2}}{\rmd %
t^{2}}\right) ^{2}}
\,.
\end{equation*}%
If we work at finite $\ve > 0$, the problem (\ref{fHD}) turns into%
\begin{eqnarray*}
\text{(HD):}\quad &&\left[ \ve ^{2}+\left( 1+\tau^{2}\frac{\rmd ^{2}}{%
\rmd t^{2}}\right) ^{2}\right] \ddot{x}=-\omega ^{2}\left( 1+\tau ^{2}%
\frac{\rmd ^{2}}{\rmd t^{2}}\right) x,\\
\text{(NL):\quad }&&
\ddot{x}=-\omega ^{2}\int_{-\infty }^{\infty }\rmd t^{\prime }\,G_{\text{f}}(t-t^{\prime })\,x(t^{\prime })\,,
\end{eqnarray*}%
and the fakeon Green function%
\begin{equation*}
G_{\text{f}}(t)=-i{\cal G}(|t|)+i{\cal G}^*(|t|),\qquad
{\cal G}(t)=\frac{\rme ^{i\sigma t/\tau }%
}{4\tau\sigma },
\qquad \sigma =\sqrt{1+i\ve },
\end{equation*}%
is straightforward to handle, with no need to extend the new NL system to a more general one.

Now the HD system admits six solutions, but four of them make the
integral on the right-hand side of the NL equation divergent. 
Thus, the solution of the NL problem contains two arbitrary constants for every $\ve >0$. We find
\begin{equation}
x_{\text{NL}}(t)=c_{1}\rme ^{i\Omega_{\ve} t}+c_{2}\rme ^{-i\Omega_{\ve}
t},\qquad c_2=c_1^*,
\label{survi0}
\end{equation}%
where $\Omega_{\ve}$ is real.

Since the reduction works for every $\ve>0$, it extends
to the limit $\ve \rightarrow 0$, which is clearly regular in  $x_{\text{NL}}(t)$. 
At the end, the NL solutions are
\begin{equation}
x_{\text{NL}}(t)=c_{1}\rme ^{i\Omega t}+c_{2}\rme ^{-i\Omega
t},\qquad \Omega =\frac{1}{\tau \sqrt{2}}\sqrt{1-\sqrt{1-4\omega^2\tau^2}},\qquad c_2=c_1^*.
\label{survi}
\end{equation}%

We have also verified the result starting from the most general NL system and showing that the fakeon limit is regular for every $\ve>0$. In particular, when cut-offs $a$ and $b$ are inserted as in
\[
\text{(NL):}\quad \ddot{x}=i\omega^2\!\int_a^t
\left[{\cal G}(t-t^\prime)-{\cal G}^*(t-t^\prime)\right]x(t^\prime)\,\textrm{d}t^\prime+i\omega^2\!\int_t^b
\left[{\cal G}(t^\prime-t)-{\cal G}^*(t^\prime-t)\right]x(t^\prime)\,\textrm{d}t^\prime ,
\]
one can check that the fakeon limit $a\rightarrow-\infty$, $b\rightarrow+\infty$ gives (\ref{survi0}) for every $\ve>0$, hence we go back to (\ref{survi}) for $\ve\rightarrow 0$.


\subsubsection{Fakeon four-derivative model}

We conclude by discussing the alternative HD case
\begin{equation}
\text{(HD):}\quad\left[ \ve ^{2}+\left( 1+\tau^{2}\frac{\rmd ^{2}}{%
\rmd t^{2}}\right) ^{2}\right] \ddot{x}=-\omega ^{2} x,
\label{HDlast}
\end{equation}%
where $\ve$ is strictly positive, otherwise the system is singular. 

The most general nonlocal problem associated with the given HD one is
\begin{eqnarray}
\text{(NL):}\quad \ddot{x}&=&-\frac{\omega^2}{\ve}\!\int_a^t
\left[{\cal G}(t-t^\prime)+{\cal G}^*(t-t^\prime)\right]x(t^\prime)\,\textrm{d}t^\prime-\frac{\omega^2}{\ve}\!\int_t^b
\left[{\cal G}(t^\prime-t)+{\cal G}^*(t^\prime-t)\right]x(t^\prime)\,\textrm{d}t^\prime\nonumber\\
&&+\int_c^d\left[A{\cal G}(t-t^\prime)+A^*{\cal G}^*(t-t^\prime)+B{\cal G}(t^\prime-t)+B^*{\cal G}^*(t^\prime-t)\right]x(t^\prime)\,\textrm{d}t^\prime\,,
\label{ridio}
\end{eqnarray}
where $A,B$ are arbitrary complex constants parametrizing the kernel of the operator appearing in the square brackets of (\ref{HDlast}).
Instead, $a,b,c$ and $d$ are introduced to ensure that the integrals are convergent.
The fakeon case is $a=-\infty$, $b=+\infty$ at $A=B=0$ (or $c=d$). 

It is easy to check that the most general six-parameter solution of the HD problem makes the integrals of the NL problem convergent, if we work directly with the fakeon prescription. This gives the impression that  all the six independent solutions survive. Nevertheless, for $a$ and $b$ generic (and $A=B=0$, for simplicity) the NL equation imposes four constraints on the six parameters of the solution, thereby reducing the number of  initial conditions of the nonlocal problem to two. 

Moreover, the fakeon limit is regular and only inherits those degrees of arbitrariness. We find the solutions
\begin{equation*}
x_{\text{NL}}(t)=c_{1}\rme ^{i\Omega t}+c_{2}\rme ^{-i\Omega
t},\qquad \Omega =\frac{\sqrt{2}}{\tau \sqrt{3}}\sqrt{1+ 2^{1/3}\Upsilon^{-1/3}-2^{-4/3}\Upsilon^{1/3}},\qquad c_2=c_1^*,
\end{equation*}%
where
\[
\Upsilon=20-27w^2+3\sqrt{3}\sqrt{16-40w^2+27w^4},\qquad w=\omega\tau\,.
\]
Note that $\Upsilon$ is positive and $\Omega$ is real.

If we rescale $\omega^2\rightarrow \ve\omega^2/\pi$ and take the
limit $\ve\rightarrow 0$, we reach the fakeonic problem
\begin{equation}
\ddot x=-\omega^2\, \delta\!\left(1+\tau^2 \frac{\textrm{d}^2}{\textrm{d}t^2}\right)x,   
\label{LDdelta}
\end{equation}
which is of the type mentioned in section \ref{sec4}.

We considered an additional variant by acting on $x$ with the operator $1+\tau^3 \textrm{d}^3/\textrm{d}t^3$ on the right-hand side of (\ref{HDlast}). The analysis proceeds as before, and the conclusion remains the same: the nonlocal problem is determined by two initial conditions, even if the right-hand side of (\ref{HDlast}) now involves three derivatives of $x$. The reason is that the left-hand side has more derivatives than the right-hand side, so perturbativity in $\omega$ is preserved.

\subsection{General proof}\label{general}

Generically, 
the presence of an infinite number of derivatives in nonlocal equations of motion may require the knowledge of infinitely many initial conditions $\varphi^{(n)}(t_{\rm i},\bm{x})$, where the superscript denotes the number of time derivatives and $t_{\rm i}$ is the initial time. This is tantamount to already knowing the solution $\varphi(t,\bm{x})$ through its Taylor expansion. In practice, it is impossible to predict the evolution of such a system without knowing it already.

Yet, the nonlocal equations of motion we are studying here, which originate from fakeons, are not of a generic type. They are intimately tied to parent higher-derivative, local equations. Hence, their set of solutions $s_{\textrm{NL}}$ is contained into the set $s_{\textrm{HD}}$  of solutions of the HD equations, which is under control. Besides, the nonlocalities are restricted to the interaction sector.

In the examples we have illustrated, the degrees of freedom of $s_{\textrm{NL}}$ coincide with those of the local equation obtained by switching off the nonlocal part. 
This suggests that in fakeon problems the nonlocal sector does not affect the number of initial conditions. Now we show that it is indeed so, by generalizing the argument of the previous subsections to generic interacting models.
For definiteness, we work with systems of the form (\ref{g1}) and (\ref{nloc}), but the result is general, as is evident from the derivation.

Consider the equation
\begin{equation}
\frac{1}{M^2}(\Box+m^2)(\Box^2+M^4)\varphi-I_{\varphi}=0\,,
\label{g1}
\end{equation}
where $I_{\varphi}$ is a generic collection of local self-interaction terms.

We can write the most general solution of (\ref{g1}) as $\varphi(x,a_i)$, where $a_i$, $i=1,\ldots 6,$ are functions that parametrize the initial conditions $\varphi^{(i)}(0,\bm{x})$, $i=0,\ldots 5$, of the Cauchy problem. 

Consider the equation
\ba
0&=&(\Box+m^2)\varphi(x)-\frac{M^2}{\Box^2+M^4}I_{\varphi}(x)\nn
&=&(\Box+m^2)\varphi(x)-i\int\rmd^4y\, G(x-y,b_j)\,I_{\varphi}(y)\,,\label{g2}
\ea
where $b_j$, $j=1,\ldots 4$ are constants that parametrize the freedom to choose a prescription for the inverse of the fourth-order operator $\Box^2+M^4$.
Precisely,
\ben
G(x,b_j)=G_{\textrm{AP}}(x)+\left\{\int \frac{\rmd^3 \bm{p}}{(2\pi)^3}\,\rme^{i \bm{p}\cdot\bm{x}}\left[
b_+(\bm{p})\,\rme^{-i  x^0 \Omega}+b_-(\bm{p})\,\rme^{i  x^0 \Omega}
\right]
+\textrm{c.c.}\right\},
\een
where $G_{\textrm{AP}}$ is the Anselmi--Piva Green function (or any other reference Green function), $b_+=b_1+i  b_2$, $b_-=b_3+i  b_4$, $\Omega=\sqrt{\bm{p}^2+ i  M^2}$ and the functions $b_i(\bm{p})$ have compact support.

Equation (\ref{g1}) is equivalent to the family of equations
(\ref{g2}) parametrized by $b_j$. Hence, $\varphi(x,a_i)$ also solve (\ref{g2}) for suitable $b_j$. Said in different words, the parameters $b_j$ can be written as functions of $a_i$. 

A quick way to see this is as follows. 
Insert the functions $\varphi(x,a_i)$ into the right-hand side of (\ref{g2}). We have no guarantee that the result is zero, but we know that what we obtain is a zero mode of the operator $(\Box^2+M^4)/M^2$, because of (\ref{g1}). Let us denote it by $\psi_0$. The kernel of $(\Box^2+M^4)/M^2$ is a four-dimensional space. Generically, we need to impose four relations among the parameters $a_i$ and $b_j$ to make $\psi_0$ disappear. 

Note that this is the point where the genericity assumption is advocated: as shown above in the models studied explicitly, it may be necessary to reach special cases as limits starting from a generic setting.

Using the relations just mentioned, we can invert four functions $a_i$ (say, $a_n$ with $n=3,4,5,6$) in terms of the other two and the parameters $b_j$. Then
we can write
the most general solution of (\ref{g2}) as
\begin{equation}
\varphi[x,a_1,a_2,a_n(a_1,a_2,b_j)].
\label{varphi}
\end{equation}
This shows that for every choice $b_j$ of the prescription for the inverse operator, the equation (\ref{g2}) has a two-parameter solution
(\ref{varphi}).

To summarize, the nonlocal system of a fakeon problem is not plagued by the need of fixing infinitely many initial conditions. 
Acting with $(\Box^2+M^4)/M^2$ on 
the right-hand side of (\ref{g2}) one obtains (\ref{g1}). Hence, the solutions of (\ref{g2}) are a subspace of the six-parameter space of solutions of (\ref{g1}). Inserting them back into (\ref{g2}), only a two-parameter subspace makes the left-hand side zero. This identifies the space of solutions of the nonlocal equations (\ref{g2}). 

We can easily extend the result to the field equations of the fakeon Lagrangian (\ref{milan}),
\begin{equation}
(\Box+m^2)\varphi+V^{\prime}-\frac{g^2}{2}\varphi\frac{M^2}{(\Box+m^2)^2+M^4}\varphi^2=0\,,
\label{nloc}
\end{equation}
by relating them to those of the local higher-derivative Lagrangian (\ref{inter}), which are \Eq{loc1} and \Eq{loc2}. Let $\varphi(x,a_i)$ and $\phi(x,a_i)$ denote the solutions of the local system \Eq{loc1}--\Eq{loc2}, which depend on six parameters $a_i$. Insert $\varphi(x,a_i)$ into (\ref{nloc}) and define 
\be
\Phi(x,a_i,b_j)\coloneqq-\frac{g}{2}\frac{M^2}{(\Box+m^2)^2+M^4}\varphi^2(x,a_i)=-\frac{ig}{2}\int d^4y\,G(x-y,b_j)\varphi^2(y,a_i),
\ee
where $b_j$ are the four parameters of the inverse operator. 

Subtracting (\ref{nloc}) to (\ref{loc1}), we find that (\ref{nloc}) holds if, and only if, $\Phi-\phi=0$. Clearly, $\Phi-\phi$ is a zero mode of the operator $[(\Box+m^2)^2+M^4]/M^2$, by \Eqq{loc2}. Hence, generically speaking, setting $\Phi-\phi$ to zero amounts to impose four relations among the parameters $a_i$ and $b_j$. We conclude that fixing the prescription $b_j$ of the inverse operator appearing in (\ref{nloc}) leaves only two initial conditions, as before.


\section{Conclusions}\label{concl}

In this paper, we have investigated the classicized Lagrangians and their equations of motion in field theories with fakeons. The presence of nonlocal interactions does not affect the number of degrees of freedom, which agrees with physical expectations. In particular, the extra modes are correctly removed by turning them into fakeons.

The classicized equations of motion are derived from those of a parent higher-derivative local system. Consequently, they avoid the need to specify infinitely many initial conditions, a difficulty that arises in generic nonlocal models. The solution space of the classicized equations coincides with the physically relevant subset of solutions of the higher-derivative system.

We have illustrated the counting of  initial conditions through a number of simple, solvable models, and provided a general proof. To some extent, Dirac’s removal of runaway solutions in classical electrodynamics can be understood along similar lines.

To avoid overcounting, it is often preferable to approach the fakeon prescription as a limit or special case of a generic prescription, instead of working with it directly.

Several physical systems may be sensitive to the effects of fakeons. Aside from primordial cosmology \cite{Anselmi:2020lpp}, we mention high-energy gravitational scattering \cite{Amati:1990xe}, early-universe expansion \cite{Mukhanov:1990me,Lyth:2009imm}, preheating and reheating phases \cite{Bassett:2005xm,Bellac:2011kqa}, and cosmic phase  transitions \cite{Enqvist:1991xw,Kapusta:2006pm}.
Generically speaking, the impact may be significant in non-stationary conditions and high-curvature regimes, whereas the effects on static or stationary systems, such as ordinary black holes, are less evident. Indeed, the fakeon prescription affects the time structure of the propagator, but has a minor impact on its spatial dependence.
Another promising research direction is fractional quantum field theory \cite{Calcagni:2021aap,Calcagni:2022shb}, where the fakeon idea plays a key role in defining correlation functions \cite{BrCa,Calcagni:2025wnn}.


\section*{Acknowledgments}

G.C.\ is supported by grant PID2023-149018NB-C41 funded by the Spanish
Ministry of Science, Innovation and Universities
MCIN/AEI/10.13039/501100011033.


\end{document}